# Enhanced nonlinear optical figure-of-merit at 1550nm for silicon nanowires integrated with graphene oxide layered films

*Yuning Zhang, Jiayang Wu, Yunyi Yang, Yang Qu, Linnan Jia, Tania Moein,*

*Baohua Jia, and David J. Moss*

Y. Zhang, Dr. J. Wu, Y. Qu, L. Jia, Dr. T. Moein, and Prof. D. J. Moss
Optical Sciences Centre
Swinburne University of Technology
Hawthorn, VIC 3122, Australia
E-mail: bjia@ swin.edu.au, dmoss@swin.edu.au

Dr. Y. Yang, Prof. B. Jia
Centre for Translational Atomaterials
Swinburne University of Technology
Hawthorn, VIC 3122, Australia

*Abstract*

Layered two-dimensional (2D) GO films are integrated with silicon-on-insulator (SOI) nanowire waveguides to experimentally demonstrate an enhanced Kerr nonlinearity, observed through self-phase modulation (SPM). The GO films are integrated with SOI nanowires using a large-area, transfer-free, layer-by-layer coating method that yields precise control of the film thickness. The film placement and coating length are controlled by opening windows in the silica cladding of the SOI nanowires. Owing to the strong mode overlap between the SOI nanowires and the highly nonlinear GO films, the Kerr nonlinearity of the hybrid waveguides is significantly enhanced. Detailed SPM measurements using picosecond optical pulses show significant spectral broadening enhancement for SOI nanowires coated with 2.2-mm-long films of 1−3 layers of GO, and 0.4-mm-long films with 5−20 layers of GO. By fitting the experimental results with theory, the dependence of GO's $n_2$ on layer number and pulse energy is obtained, showing interesting physical insights and trends of the layered GO films from 2D monolayers to quasi bulk-like behavior. Finally, we show that by coating SOI nanowires with GO films the effective nonlinear parameter of SOI nanowires is increased 16 fold, with the effective nonlinear figure of merit (FOM) increasing by about 20 times to FOM > 5. These results reveal the strong potential of using layered GO films to improve the Kerr nonlinear optical performance of silicon photonic devices.

# 1. Introduction

The 3$^{rd}$ order nonlinear optical response has found wide ranging applications in telecommunications, metrology, astronomy, ultrafast optics, quantum photonics, and many other areas [1-4]. Nonlinear integrated photonic devices based on the Kerr ($n_2$) effect in particular offer far superior processing speeds compared to electronic devices as well as the added benefits of compact footprint, low power consumption, high stability, and low-cost mass production, all of which are important for high-speed signal generation and processing in optical communication systems [5-7].

As a leading platform for integrated photonic devices, silicon-on-insulator (SOI) leverages the well-developed complementary metal-oxide-semiconductor (CMOS) fabrication technology for the implementation of functional devices for optical communications [8-10], optical interconnects [11-13], photonic processing [14-16], and biosensing [17, 18]. Nevertheless, the strong two-photon absorption (TPA) of silicon at near-infrared wavelengths poses a fundamental limitation to the performance of nonlinear photonic devices in the telecommunications band [5, 6]. Other CMOS compatible platforms such as silicon nitride [19, 20], high index doped silica glass [21, 22], and silicon rich nitride [23, 24], have a much weaker TPA and a higher nonlinear figure of merit (FOM), although they still face limitations in terms of nonlinear efficiency due to their lower intrinsic Kerr nonlinearity [5, 25].

The need to improve the performance of nonlinear integrated photonic devices has motivated the on-chip integration of highly nonlinear materials such as polymers and two-dimensional (2D) materials [26, 27]. The giant Kerr nonlinearity of 2D layered materials such as graphene, graphene oxide (GO), black phosphorus, and transition metal dichalcogenides (TMDCs) has been widely recognized and has enabled diverse nonlinear photonic devices with high performance and new functionalities [28-32]. In particular, enhanced spectral broadening of optical pulses has been reported for SOI nanowires with transferred MoS$_2$ and graphene [33-35].

Among the various 2D materials, GO has received increasing interest due to its ease of preparation as well as the tunability of its material properties [36-41]. Previously, we reported GO films with a giant Kerr nonlinearity of about 4 orders of magnitude higher than that of silicon [40, 42], and demonstrated significantly enhanced four-wave mixing (FWM) in doped silica waveguides and microring resonators integrated with layered GO films [43, 44]. Moreover, GO has a material absorption that is over 2 orders of magnitude lower than undoped graphene as well as a large bandgap (2.1−2.4 eV) that yields a low TPA in the telecommunications band [45, 46]. By using a large-area, transfer-free, layer-by-layer GO coating method, we also achieved highly precise control of the GO film thickness on integrated devices [45, 47]. This overcomes critical fabrication limitations in terms of layer transfer for 2D materials and represents a significant improvement for the eventual manufacturing of integrated photonic devices incorporated with 2D layered GO films.

In this paper, we use our GO fabrication technique to demonstrate a significantly enhanced Kerr nonlinearity in silicon-on-insulator (SOI) nanowires integrated with 2D layered GO films. Self-phase modulation (SPM) measurements are performed at different pulse energies for the SOI nanowires integrated with different numbers of GO layers (2.2-mm-long with 1−3 layers of GO and 0.4-mm-long with 5−20 layers of GO). Benefiting from the strong light-matter interaction between the SOI nanowires and the highly nonlinear GO films, we observe significant spectral broadening for the GO-coated SOI nanowires as compared with the uncoated SOI nanowires, achieving a high broadening factor (BF) of 3.75 for an SOI nanowire with 2 layers of GO and 4.34 for a device with 10 layers of GO. We also fit the SPM experimental results to theory and obtain the dependence of the Kerr nonlinearity of the GO films on the GO layer number and pulse energy, showing interesting physical insights and trends of the layered GO films in evolving from 2D monolayers to quasi bulk-like behavior. Finally, we obtain the effective nonlinearity ($n_2$), nonlinear parameter ($γ$) and nonlinear FOM of the hybrid waveguides, showing that the GO films can enhance $γ$ of SOI nanowires by up to 16 folds and the nonlinear FOM by almost 20 times. These results verify the effectiveness of integrating 2D layered GO films with silicon photonic devices to improve the performance of Kerr nonlinear optical processes.

## 2. Device fabrication and characterization

*2.1 Device fabrication*

**Figure 1(a)** shows a schematic of an SOI nanowire waveguide integrated with a GO film while the fabrication process flow is given in **Figure 1(b)**. SOI nanowires with a cross section of 500 nm × 220 nm were fabricated on an SOI wafer with a 220-nm-thick top silicon layer and a 2-μm-thick buried oxide (BOX) layer via CMOS-compatible fabrication processes. Deep ultraviolet photolithography (248 nm) was used to define the device layout, followed by inductively coupled plasma etching of the top silicon layer. After that, a 1.5-μm thick silica layer was deposited by plasma enhanced chemical vapor deposition as an upper cladding layer. Windows with two different lengths of 0.4 mm and 2.2 mm were then opened down to the BOX layer via photolithography and reactive ion etching, so as to enable GO film coating onto the SOI nanowire. Finally, the coating of 2D layered GO films was achieved by a solution-based method that yielded layer-by-layer GO film deposition, as reported previously [43, 45, 47]. Four steps for the in-situ assembly of monolayer GO films were repeated to construct multilayer films on the SOI nanowire. Our GO coating approach, unlike the sophisticated transfer processes employed for coating other 2D materials such as graphene and TMDCs [33, 34], enables transfer-free GO film coating on integrated photonic devices, with highly scalable fabrication processes and precise control of the number of GO layers (i.e., GO film thickness).

**Figure 2(a)** shows a microscope image of a fabricated SOI chip with a 0.4-mm-long opened window. Apart from allowing precise control of the placement and coating length of the GO films that are in contact with the SOI nanowires, the opened windows also enabled us to test the performance of devices having a shorter length of GO film but with higher film thicknesses (up to 20 layers). This provided more flexibility to optimize the device performance with respect to SPM spectral broadening.

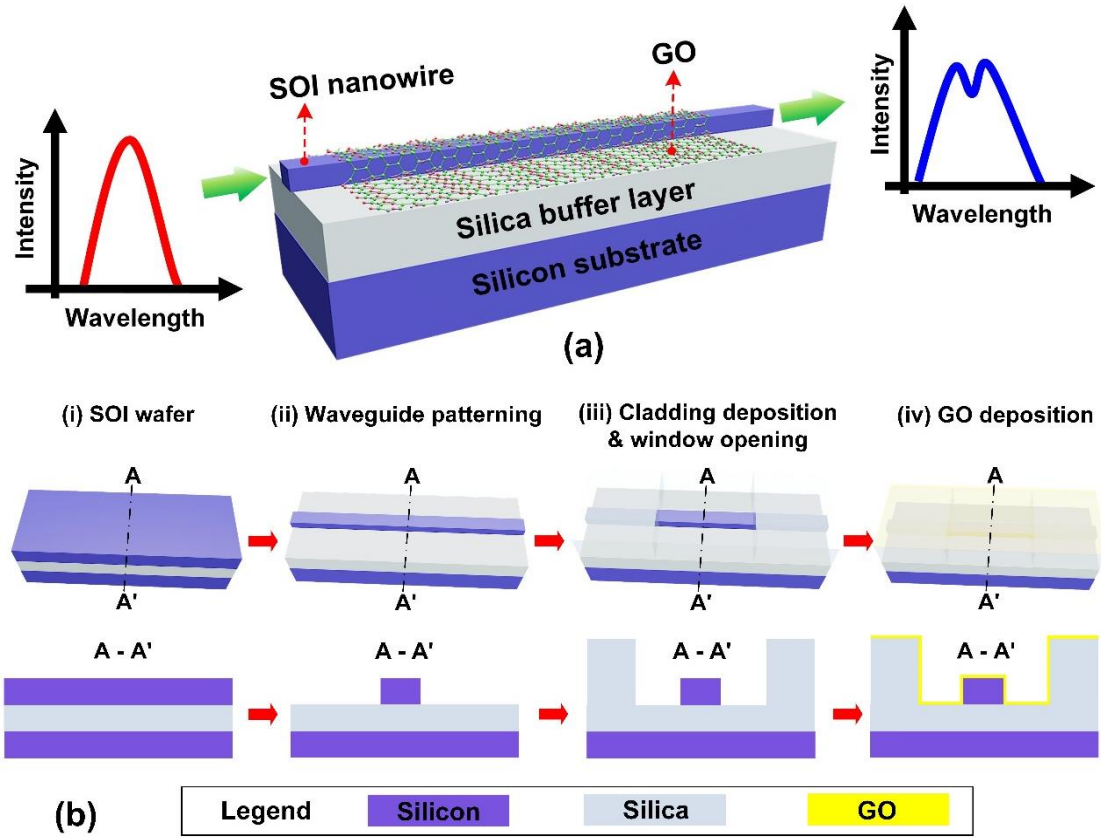

**Figure 1**. (a) Schematic illustration of a GO-coated SOI nanowire waveguide. (b) Schematic illustration showing the fabrication process flow.

**Figure 2(b)** shows the scanning electron microscopy (SEM) image of an SOI nanowire conformally coated with 1 layer of GO. Note that the conformal coating (with the GO film coated on both the top surface and sidewalls of the nanowire) is slightly different to earlier work where we reported doped silica devices with GO films only coated on the top surface of the waveguides [43, 47]. As compared with doped silica waveguides, the SOI nanowires allow much stronger light-material interaction between the evanescent field leaking from the waveguide and the GO film, which is critical to enhance nonlinear optical processes such as SPM and FWM.

**Figure 2(c)** presents an SEM image of GO films with up to 5 layers of GO, clearly showing the layered film structure. The successful integration of GO films is confirmed by the representative D (1345 cm$^{-1}$) and G (1590 cm$^{-1}$) peaks of GO observed in the Raman spectrum of an SOI chip coated with 5 layers of GO, as shown in **Figure 2(d)**. Microscope images of the same SOI chip before and after GO coating are shown in the insets, which illustrate good morphology of the films.

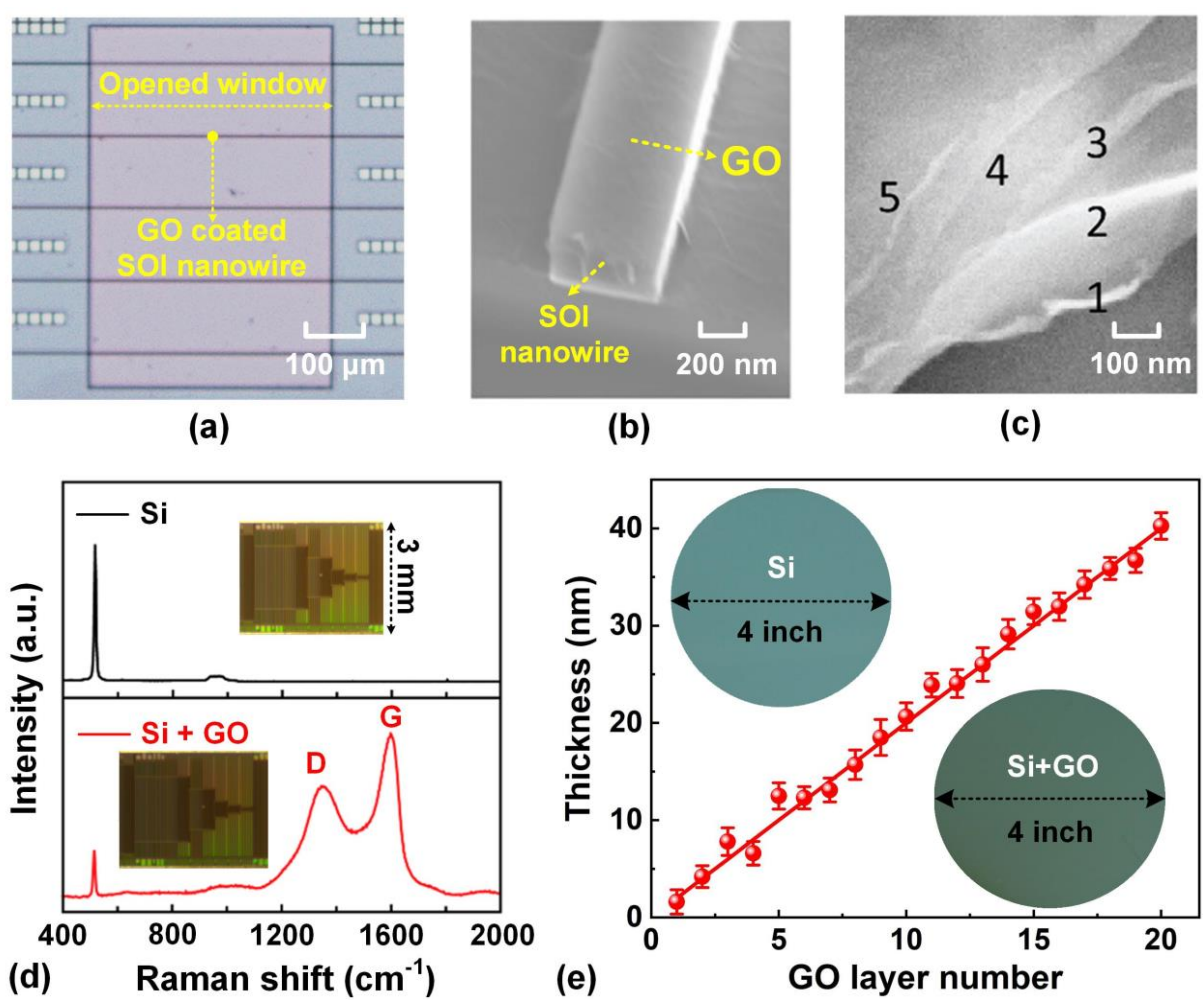

**Figure 2.** (a) Microscope image of a fabricated SOI chip with a 0.4-mm-long opened window. (b) Scanning electron microscopy (SEM) image of a SOI nanowire conformally coated with 1 layer of GO. (c) SEM image of layered GO film. The numbers refer to the number of layers for that part of the image. (d) Raman spectra of an SOI chip without GO and with 5 layers of GO. Insets show the corresponding microscope images. (e) Measured GO film thickness versus GO layer number. The plots show the average of measurements on three samples and the error bars reflect the variations. Insets show the images of an uncoated silicon wafer and the same wafer uniformly coated with 20 layers of GO.

**Figure 2(e)** shows the GO film thickness versus number of layers characterized via atomic force microscopy. The dependence of GO film thickness versus layer number shows a nearly linear relationship, with a thickness of ~2.06 nm on average for each layer. The insets show photo images of a 4-inch silicon wafer before and after being uniformly coated with 20 layers of GO. The high uniformity of the GO films demonstrates the capability for large-area uniform coating, which allows large-scale integrated devices incorporated with 2D layered GO films.

*2.2 Linear loss measurement*

We fabricated and tested two types of GO-coated SOI nanowires with (i) 2.2-mm-long, 1−3 layers of GO and (ii) 0.4-mm-long, 5−20 layers of GO. **Figures 3(a-i) and (a-ii)** depict the insertion loss of the GO-coated SOI nanowires, all measured using a low-power (0 dBm) transverse electric (TE) polarized continuous-wave (CW) light. The total length of the SOI nanowires (including the segments with and without silica cladding) was 3.0 mm. We employed lensed fibers to butt couple the CW light into and out of the SOI nanowires with inverse-taper couplers at both ends with a butt coupling loss of ~5 dB per facet. **Figure 3(b)** shows the propagation loss of the GO-coated SOI nanowires and the bare (uncoated) SOI nanowire. The latter one was ~0.43 dB/mm obtained from cutback measurements of SOI nanowires with different lengths. The propagation loss of the SOI nanowire with a monolayer of GO was ~2.48 dB/mm, corresponding to an excess propagation loss of ~2.05 dB/mm induced by the GO film. This is about 2 orders of magnitude smaller than SOI nanowires coated with graphene [48], indicating the low material absorption of GO and its strong potential for the implementation of high-performance nonlinear photonic devices. The propagation loss increased with GO layer number − a combined result of increased mode overlap and several other possible effects such as increased scattering loss and absorption induced by imperfect contact between the multiple GO layers as well as interaction between the GO layers, as reported previously [43, 47].

Unlike graphene that has a metallic behavior (e.g., high electrical and thermal conductivity) with a zero bandgap, GO is a dielectric that has a large bandgap of 2.1−2.4 eV [36, 45], which results in low linear light absorption in spectral regions below the bandgap. In theory, GO films with a bandgap > 2 eV should have negligible absorption for light at near-infrared wavelengths. We therefore infer that the linear loss of the GO films is mainly due to light absorption from localized defects as well as scattering loss stemming from film unevenness and imperfect contact between the different layers.

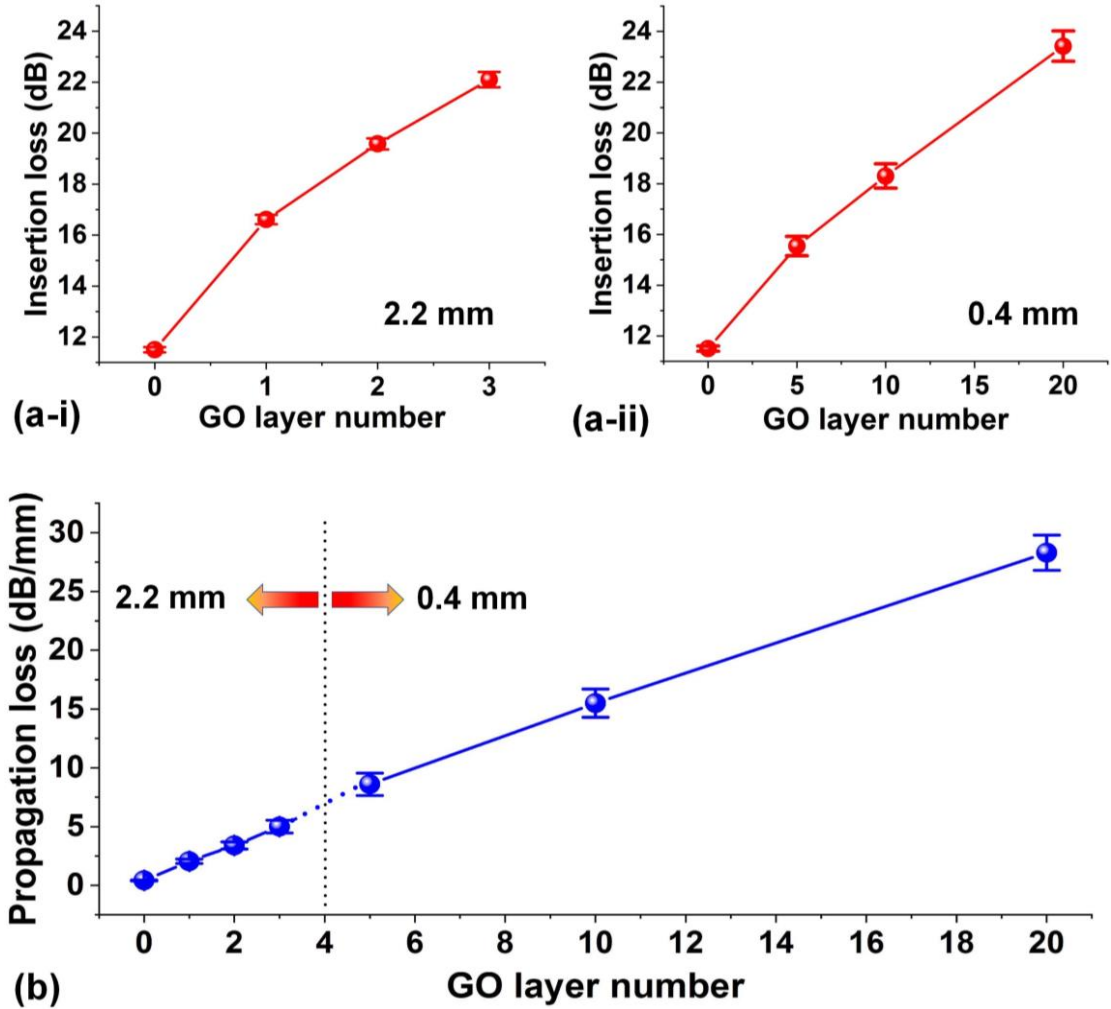

**Figure 3.** Linear loss measurement. (a) Measured insertion loss of SOI nanowires with (i) 2.2-mm-long, 1−3 layers of GO and (ii) 0.4-mm-long, 5−20 layers of GO. (b) Propagation loss of the GO-coated SOI nanowires extracted from (a-i) and (a-ii). In (a) and (b), the results for the bare SOI nanowires (i.e., the GO layer number is 0) are also shown for comparison. The data points depict the average values obtained from the measured insertion loss of three duplicate devices and the error bars illustrate the variations among the different samples.

*2.3 Nonlinear loss measurement*

We measured the nonlinear loss of the GO-coated SOI nanowires using a pulsed fiber laser (PriTel, repetition rate: ~60 MHz, pulse duration: ~3.9 ps). The experimental setup is shown in **Figure 4(a)**, where two optical power meters (OPM 1 and OPM 2) were employed to measure the average power of the optical pulses before and after transmission through the SOI nanowires. A variable optical attenuator (VOA) was used to tune the input pulse energy. **Figures 4(b-i)** and **4(b-ii)** depict the measured power-dependent excess insertion loss (EIL, defined as the additional insertion loss over that of **Figures 3(a-i) and 3(a-ii)**, respectively) as a function of the coupled pulse energy (0.38−51.5 pJ, i.e., 0.1−13.2 W coupled peak power, excluding the butt coupling loss) for the SOI nanowires with (i) 2.2-mm-long, 1−3 layers of GO and (ii) 0.4-mm-long, 5−20 layers of GO. The corresponding results for the bare SOI nanowires are also shown for comparison, where the EIL increased with pulse energy primarily due to TPA and free carrier absorption of silicon [49]. For the GO-coated SOI nanowires, on the other hand, the measured EIL was slightly lower than that for the bare SOI nanowire, with the difference increasing with GO layer number. This reflects the power-dependent loss of the GO layers (see Section 4 for detailed discussion). **Figure 4(c)** shows the ΔEIL for the hybrid waveguides extracted from **Figure 4(b)** after excluding the EIL induced by the bare SOI nanowire, which reflects the power dependent loss due solely to the GO films. The ΔEIL decreased as the pulse energy was increased – a trend that is consistent with saturable absorption (SA) [50, 51] (see Section 4 for detailed discussion). We also note that these changes were not permanent – the measured insertion loss recovered to those in **Figures 3(a-i)** and **(a-ii)** when the pulse energy was reduced, with the measured EIL in **Figures 4(b-i)** and **(b-ii)** being repeatable.

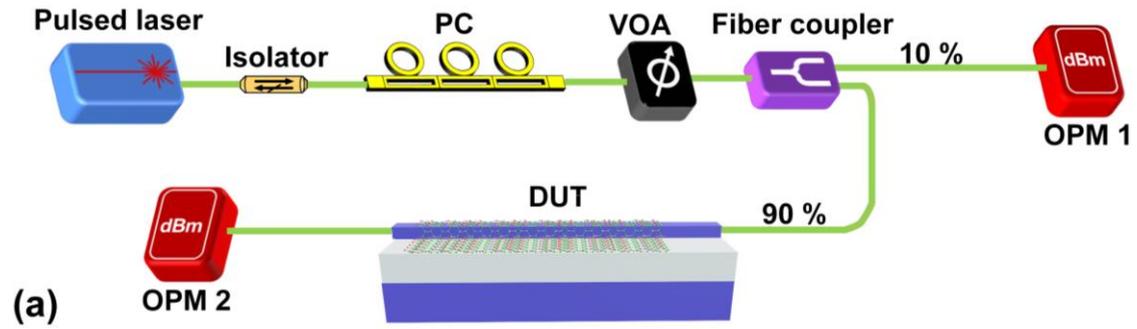

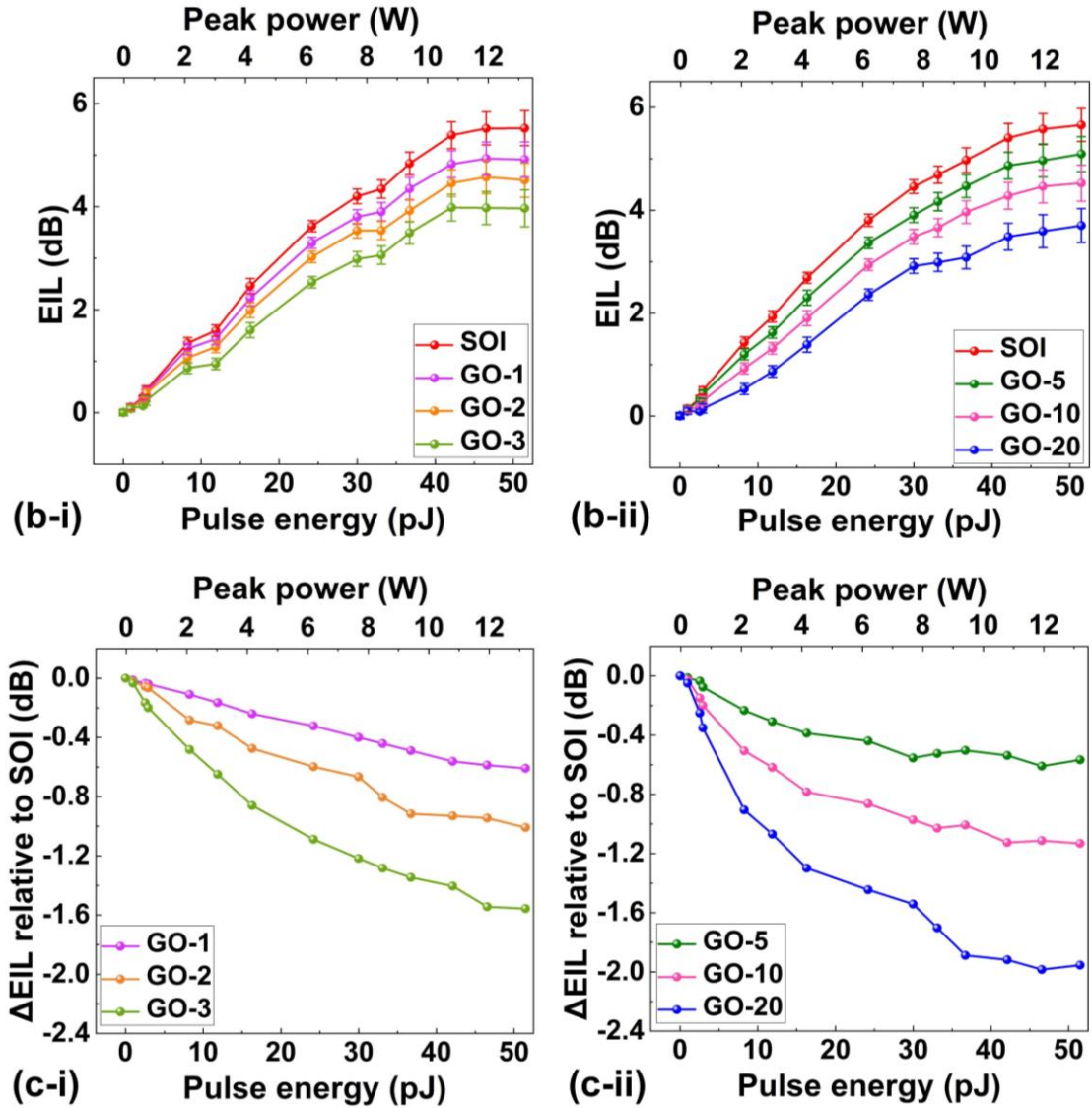

**Figure 4.** Nonlinear loss measurement. (a) Experimental setup for measuring nonlinear loss of the GO-coated SOI nanowires. PC: polarization controller. VOA: variable optical attenuator. OPM: optical power meter. DUT: device under test. (b) Power-dependent excess insertion loss (EIL) of optical pulses after transmission through SOI nanowires with (i) 2.2-mm-long, 1−3 layers of GO and (ii) 0.4-mm-long, 5−20 layers of GO. The results for the bare SOI nanowires (SOI) are also shown for comparison. (c) ΔEIL relative to SOI extracted from (b).

## 3. SPM experiments

We used the experimental setup shown in **Figure 5** to perform SPM measurements on the GO-coated SOI nanowires. Picosecond optical pulses generated by a pulsed fiber laser (the same as in **Section 2.3**) were delivered into the hybrid waveguides, with a VOA to tune the input pulse energy. An optical isolator was inserted before the device under test (DUT) to prevent the reflected light from damaging the laser source and a polarization controller (PC) was used to set the input light to TE-polarization. The signal output from the DUT was split by a 90:10 coupler – 10% sent into an optical power meter (OPM) for power monitoring and the other 90% into an optical spectrum analyzer (OSA, Yokogawa AQ6370D) to observe the spectral broadening.

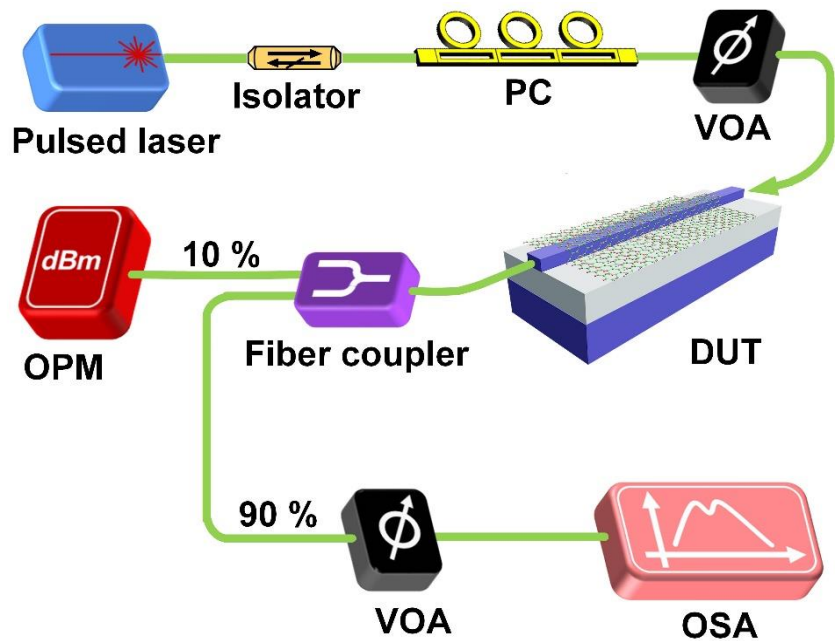

**Figure 5.** Experimental setup for SPM measurement in GO-coated SOI nanowires. PC: polarization controller. VOA: variable optical attenuator. OPM: optical power meter. DUT: device under test. OSA: optical spectrum analyzer.

**Figure 6** shows the results of the SPM experiments. **Figure 6(a-i)** shows the normalized spectra of the optical pulses before and after transmission through the SOI nanowires with 2.2-mm-long, 1–3 layers of GO, together with the output optical spectrum for the bare SOI nanowire, all taken with the same pulse energy of ~51.5 pJ (i.e., ~13.2 W peak power, excluding coupling loss) coupled into the SOI nanowires. As compared with the input pulse spectrum, the output spectrum after transmission through the bare SOI nanowire exhibited measurable spectral broadening. This is expected and can be attributed to the high Kerr nonlinearity of silicon. The GO-coated SOI nanowires, on the other hand, show much more significantly broadened spectra as compared with the bare SOI nanowire, clearly reflecting the improved Kerr nonlinearity of the hybrid waveguides. **Figure 6(a-ii)** shows the corresponding results for the SOI nanowires with 0.4-mm-long, 5–20 layers of GO, taken with the same coupled pulse energy as in **Figure 6(a-i)**. The SOI nanowires with a shorter GO coating length but higher film thicknesses also clearly show more significant spectral broadening as compared with the bare SOI nanowire. We also note that in either **Figure 6(a-i)** or **6(a-ii)**, the maximum spectral broadening is achieved for a device with an intermediate number of GO layers (i.e., 2 and 10 layers of GO in **(a-i)** and **(a-ii)**, respectively). This could reflect the trade-off between the Kerr nonlinearity enhancement (which dominates for the device with a relatively short GO coating length) and loss increase (which dominates for the device with a relatively long GO coating length) for the SOI nanowires with different numbers of GO layers (see Section 4 for detailed discussion).

**Figures 6(b-i)** and **(b-ii)** show the power-dependent output spectra after going through the SOI nanowires with (i) 2 layers and (ii) 10 layers of GO films. We measured the output spectra at 10 different coupled pulse energies ranging from ~8.2 pJ to ~51.5 pJ (i.e., coupled peak power from ~2.1 W to ~13.2 W). As the coupled pulse energy was increased, the output spectra showed increasing spectral broadening as expected. We also note that the broadened spectra exhibited a slight asymmetry. This was a combined result of both the asymmetry of the input pulse spectrum and the free-carrier dispersion of silicon [49] (see Section 4).

To quantitatively analyze the spectral broadening of the output spectra, we introduce the concept of a broadening factor (BF) [52, 53]. The width of the optical spectra is described by the root-mean-square (RMS) width, defined as [52]

$$\Delta\omega_{rms}^2=\langle(\omega-\omega_0)^2\rangle-\langle(\omega-\omega_0)\rangle^2 \tag{1}$$

where the angle brackets denote the average over the spectrum given by

$$\begin{cases} \langle(\omega-\omega_0)^2\rangle = \frac{\int_{-\infty}^{\infty}(\omega-\omega_0)^2 S(\omega)d\omega}{\int_{-\infty}^{\infty}S(\omega)d\omega} \\ \langle(\omega-\omega_0)\rangle^2 = [\frac{\int_{-\infty}^{\infty}(\omega-\omega_0)S(\omega)d\omega}{\int_{-\infty}^{\infty}S(\omega)d\omega}]^2 \end{cases} \tag{2}$$

where $S(\omega)$ is the spectrum intensity. The BF is therefore defined as:

$$BF = \frac{\Delta\omega_{rms}}{\Delta\omega_0} \tag{3}$$

where $\Delta\omega_0$ is the RMS spectral width of the input optical pulses.

**Figure 6(c)** shows the BFs of the measured output spectra after transmission through through the bare SOI nanowire and the GO-coated SOI nanowires at pulse energies of 8.2 pJ and 51.5 pJ. The GO-coated SOI nanowires show higher BFs than the bare SOI nanowires (i.e., GO layer number = 0), and the BFs at a coupled pulse energy of 51.5 pJ are higher than those at 8.2 pJ, agreeing with the results in **Figure 6(a)** and **6(b)**, respectively. At 51.5 pJ, BFs of up to 3.75 and 4.34 are achieved for the SOI nanowires with 2 and 10 layers of GO, respectively. This also agrees with the results in **Figure 6(a)** − with the maximum spectral broadening being achieved for an intermediate number of GO layers due to the trade-off between the Kerr nonlinearity enhancement and increase in loss. The BFs of the output spectra versus coupled pulse energy are shown in **Figures 6(d-i)** and **6(d-ii)** for the SOI nanowires with 1−3 layers and 5−20 layers of GO, respectively. The BFs increase with coupled pulse energy, reflecting a more significant spectral broadening that agrees with the results in **Figure 6(b)**.

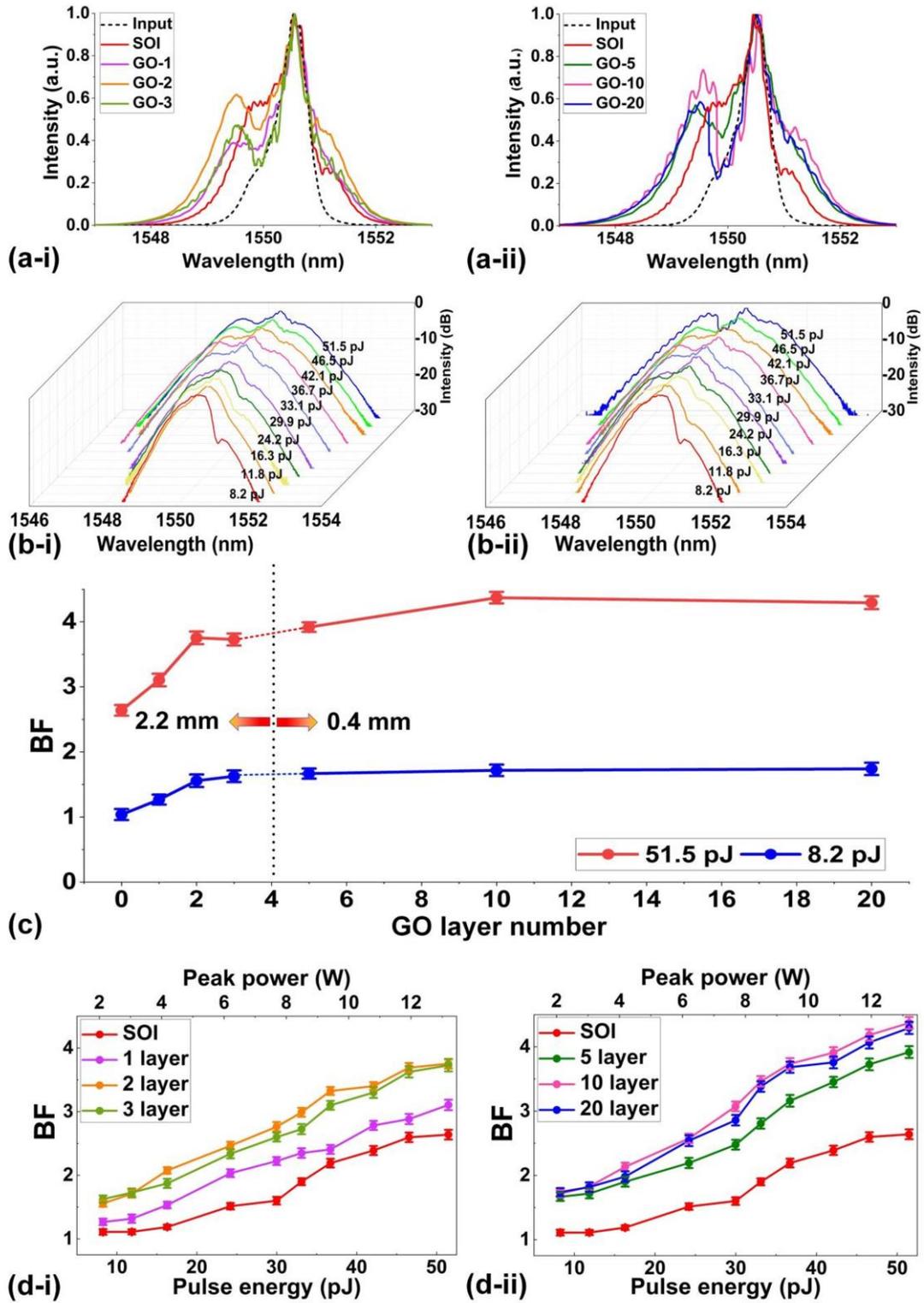

**Figure 6.** SPM experimental results. (a) Normalized spectra of optical pulses before and after going through the GO-coated SOI nanowires at a coupled pulse energy of ~51.5 pJ. (b) Optical spectra measured at different pulse energies for the GO-coated SOI nanowires. (c) BFs of the measured output spectra versus GO layer number at fixed coupled pulse energies of 8.2 pJ and 51.5 pJ. (d) BFs of the measured output spectra versus coupled pulse energy (or coupled peak power). In (a), (b) and (d), (i) and (ii) show the results for the SOI nanowires with 2.2-mm-long, 1−3 layers of GO and with 0.4-mm-long, 5−20 layers of GO, respectively. In (a), (c) and (d), the corresponding results for the bare SOI nanowires are also shown for comparison.

## 4. Theoretical analysis and discussion

*4.1 SPM spectral broadening*

We used the theory from Refs. [49, 50, 52] to model the SPM process in the GO-coated SOI nanowires. The evolution of an optical pulse going through the hybrid waveguides was simulated by using a split-step Fourier method to solve the nonlinear Schrödinger equation (NLSE) as follows [49]:

$$\frac{\partial A}{\partial z} = -\frac{i\beta_2}{2}\frac{\partial^2 A}{\partial t^2} + i\gamma |A|^2 A - \frac{1}{2}i\sigma\mu N_c A - \frac{1}{2}\alpha A \quad (4)$$

where $i = \sqrt{1}$, $A(z, t)$ is the slowly varying temporal envelope of the optical pulse along the z axis, which is the direction of light propagation, $\beta_2$ is the second-order dispersion coefficient, $\gamma$ is the waveguide nonlinear parameter, $\sigma$ and $\mu$ are the free carrier absorption (FCA) and free carrier dispersion (FCD) coefficients of silicon, respectively, $N_c$ is the free carrier density in silicon, and $\alpha$ is the total loss including both linear loss and nonlinear loss.

In **Eq. (4)**, we retain only the second-order dispersion $\beta_2$ since the dispersion length (> 1 m) is much longer than the physical length of the waveguides (i.e., the third-order nonlinearity dominates during the pulse propagation). The total loss $\alpha$ can be expressed as:

$$\alpha = \alpha_L + \alpha_{NL\text{-}SOI} + \alpha_{NL\text{-}GO} \quad (5)$$

where $\alpha_L$ is the linear loss, $\alpha_{NL\text{-}SOI}$ and $\alpha_{NL\text{-}GO}$ are the nonlinear losses induced by the SOI nanowire and the GO film, respectively.

In **Eq. (5)**, $\alpha_{NL\text{-}SOI}$ includes the nonlinear losses induced by TPA and FCA of silicon, which are [49]:

$$\alpha_{NL\text{-}SOI} = \frac{\beta_{TPA,Si}}{A_{eff}}|A|^2 + \sigma N_c \quad (6)$$

where $\beta_{TPA,Si}$ is the TPA coefficient of silicon and $A_{eff}$ is the effective mode area of the waveguides. In **Eqs. (4)** and **(6)**, $N_c$ results from free carriers generated in silicon, given by [49]:

$$\frac{\partial N_c(z,t)}{\partial t} = \frac{\beta_{TPA,Si}}{2\hbar\omega} \cdot \frac{|A(z,t)|^4}{A_{eff}^2} - \frac{N_c(z,t)}{\tau_c} \quad (7)$$

where $\hbar$ is the Planck's constant, $\omega$ is the angular frequency, $\tau_c = \sim 1$ ns is the effective carrier lifetime. The $\tau_c$ term in **Eq. (7)** can be ignored since the pulse duration (~3.9 ps) is much less than $\tau_c$ so that the generated free carriers do not have time to recombine within the pulse duration [34, 49]. The measured EIL of the bare SOI nanowire as a function of the coupled pulse energy is shown in **Figure 7(a)** along with the fit EIL calculated from **Eqs. (6)−(7)**. The fit $\beta_{TPA, Si}$ and $\sigma$ are 5.02 ×10$^{-12}$ m/W and 1.44×10$^{-21}$ m$^2$, respectively, which agree well with the literature [49], confirming that the nonlinear loss of the bare SOI nanowire was mainly induced by TPA and FCA. We also note that FCD significantly affected the output spectral shapes, largely accounting for the significant asymmetry.

Since the TPA and FCA of the GO films is negligible in the telecommunications band [42, 43], $\alpha_{NL-GO}$ in **Eq. (5)** is mainly a result of SA in the GO films, as noted previously [40, 42, 51]. In **Figure 7 (b)**, we exclude the nonlinear loss induced by the SOI nanowires before the GO coated segments (0.4 mm for the devices with 2.2-mm-long GO films and 1.3-mm for the devices with 0.4-mm-long GO films) from the total pulse energy coupled into the hybrid waveguides and obtain the pulse energy at the start of the GO coated segments, which is the pulse energy for SA. Based on the theory in [54, 55], the SA loss of the GO films was modelled via the following equation:

$$\alpha_{NL-GO} = \alpha_{sat} /(1 + \frac{|A|^2}{I_s}) \tag{8}$$

where $\alpha_{sat}$ is the saturable absorption coefficient and $I_s$ is the saturation intensity. The measured and fit ΔEIL of the hybrid waveguides with 1−3 layers and 5−20 layers of GO are depicted in **Figures 7 (c-i)** and **(c-ii)**, respectively. The fit curves were calculated with **Eq. (8),** and ΔEIL is replotted as a function of pulse energy at the start of the GO coated segments, versus the total pulse energy coupled into the hybrid waveguides in **Figure 4(c)**. **Figure 7(d)** shows the fit $\alpha_{sat}$ and $I_s$ for different GO layer numbers, showing that $\alpha_{sat}$ and $I_s$ are not constant. This reflects the change in the nonlinear properties of 2D GO films with layer number − $\alpha_{sat}$ increases with the GO layer number, whereas $I_s$ shows the opposite trend. This reflects an increased SA efficiency and decreased power threshold for thicker GO films. At low powers where the SA of GO is negligible, the linear loss in **Figure 3(b)** corresponds to $α_L + α_{sat}$. At high light powers where the SA of GO cannot be neglected, on the other hand, the measured ΔEIL in **Figure 4(c)** corresponds to $α_{NL\text{-}GO} - α_{sat}$. The SA of GO is mainly induced by the ground-state bleaching of the $sp^2$ related states [40] that have a narrow energy gap of ~0.5 eV, where the optical absorption of electrons can be easily saturated, depleting the valence band and filling the conduction band [56, 57]. Note that the spectral broadening induced by SPM has negligible direct impact on the saturable absorption since the absorption induced by GO was wavelength independent in our model due to the broadband response of the 2D layered GO films [47]. In our calculations, we neglected any potential spectral/temporal coupling effects that could induce interplay between SA and SPM. This was because the length of the hybrid SOI nanowires was much shorter than the dispersion length for the picosecond pulses used, and so the change in the pulse spectrum due to SPM did not have a significant effect on the temporal shape of the pulse [50]. According to our simulations, the maximum difference induced by SPM on ΔEIL is < 1%.

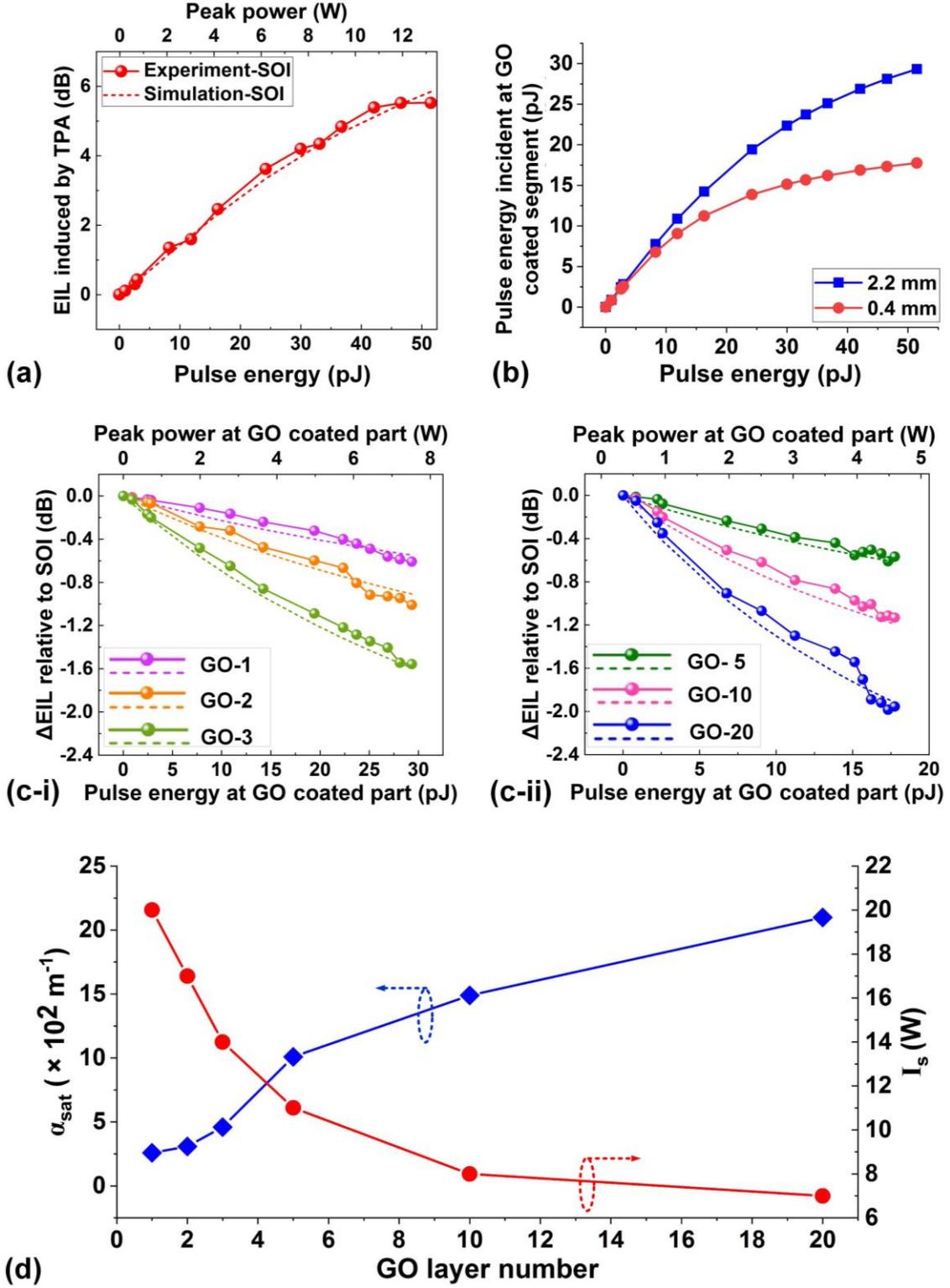

**Figure 7.** (a) Measured and simulated EIL induced by TPA and FCA of silicon. (b) Pulse energy incident at the 2.2-mm-long and 0.4-mm-long GO coated segments versus pulse energy coupled into the hybrid waveguides. (c) Measured (solid curves) and fit (dashed curves) ΔEIL relative to SOI versus pulse energy incident at (i) 2.2-mm-long and (ii) 0.4-mm-long GO coated segments. (d) Fit $α_{sat}$ and $I_s$ versus GO layer number.

Based on **Eqs. (4)−(8)**, we fit the experimentally measured spectra to obtain the effective nonlinear parameter $γ_{eff}$ of the hybrid waveguides. In our calculations, the hybrid waveguides were divided into SOI (with silica cladding) and hybrid (with GO in opened windows) segments and the SPM differential equation in **Eq. (4)** was solved for each segment. **Figure 8(a)** shows the measured and fit optical spectra for the input pulses. We approximated the input pulse shape by a Gaussian profile as below:

$$A=\sqrt{P_0} \cdot exp[-\frac{1}{2}(1+iC_0)(\frac{t}{T_0})^2] \qquad (9)$$

where $P_0$ is the pulse peak power, $C_0$ is the initial chirp, and $T_0$ is se duration. The measured and fit optical spectra for the output signal after transmission through the bare SOI nanowire at a coupled pulse energy of 51.5 pJ are shown in **Figure 8(b)**, showing good agreement between theory and experiment, with the discrepancy arising mainly from small imperfections in the input pulse spectrum. We obtain a $\gamma$ of 288 $W^{-1}m^{-1}$ for the bare SOI nanowire, in agreement with previous reports [6]. The measured and fit optical spectra for the output signals after transmission through the SOI nanowires with 2 and 10 layers of GO are shown in **Figures 8(c)** and **(d)**, respectively, where the coupled pulse energy is the same as **Figure 8(b)**. Here again, we obtain good agreement between theory and experiment, particularly for the BFs where the deviation is < 3.2 %. This small discrepancy is mainly due to slight differences between the measured and fit output spectra. **Figure 8(e)** shows the effective nonlinear parameter of the hybrid waveguides ($\gamma_{eff}$, nomalized to $\gamma$ of the bare SOI nanowire) as a functional of GO layer number. As expected, $\gamma_{eff}$ increases with the GO layer number − a combined result of the change of mode overlap as well as $n_2$ of the GO films with layer number. In particular, for the SOI nanowires with 10 or more layers of GO, $\gamma_{eff}$ is over 10 times that of the bare SOI nanowire. In our calculations we neglected any slight change of mode overlap due to the change in film thickness and refractive index of GO with pulse energy during SPM.

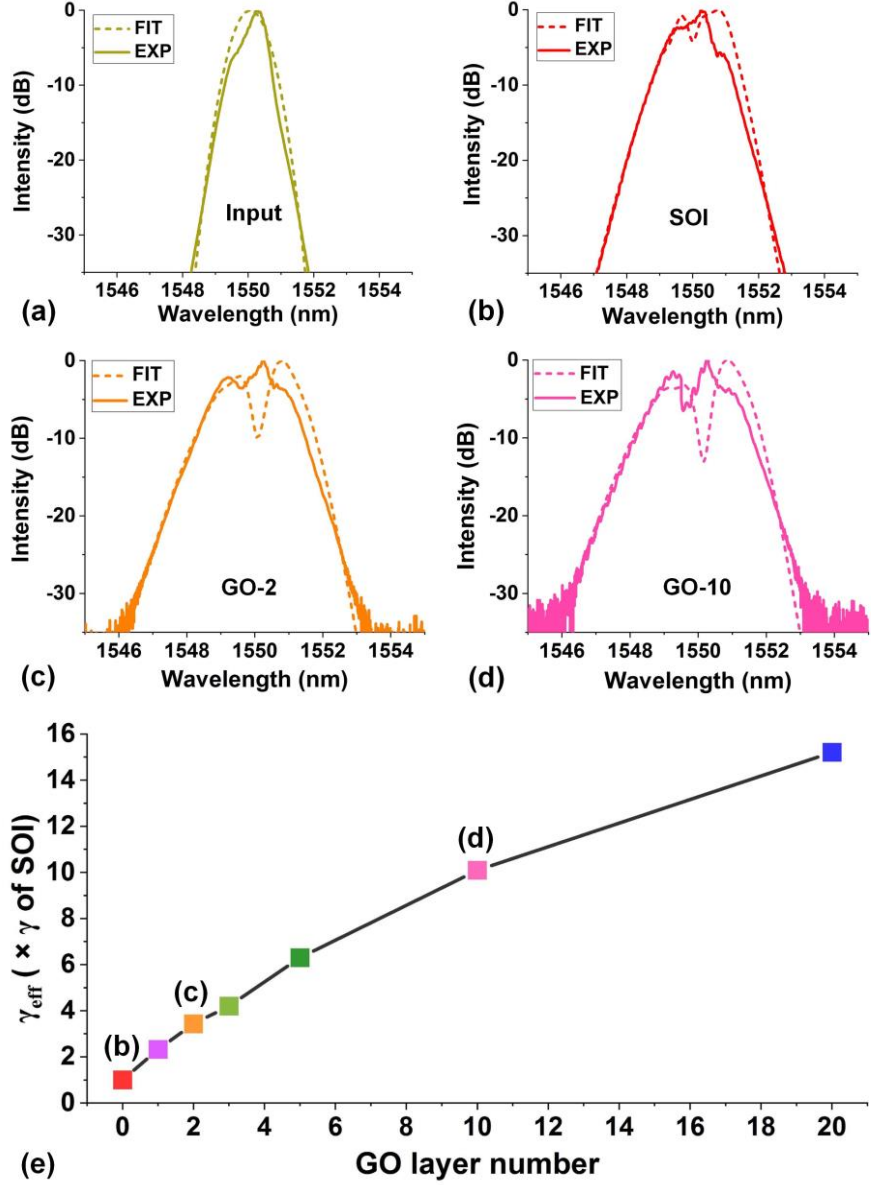

**Figure 8.** (a)−(d) Measured and fit optical spectra for (a) input optical pulses, (b) output signal after transmission through the bare SOI nanowire, (c) output signal after transmission through the SOI nanowire with 2.2-mm-long, 2 layers of GO, and (d) output signal after transmission through the SOI nanowire with 0.4-mm-long, 10 layers of GO. (e) $\gamma_{eff}$ of hybrid waveguides (normalized to $\gamma$ of the bare SOI nanowire) versus GO layer number. The data points corresponding to (b)−(d) are labelled in (e). The coupled pulse energy in (b)−(e) is ~51.5 pJ.

## 4.2 Kerr nonlinearity ($n_2$) of the GO films

Based on $\gamma_{eff}$ of the hybrid waveguides, we further extract the Kerr coefficient ($n_2$) of the layered GO films using [43]:

$$\gamma_{eff} = \frac{2\pi}{\lambda_c} \frac{\iint_D n_0^2(x,y) n_2(x,y) S_z^2 dxdy}{\left[\iint_D n_0(x,y) S_z dxdy\right]^2} \qquad (10)$$

where $\lambda_c$ is the pulse central wavelength, $D$ is the integral of the optical fields over the material regions, $S_z$ is the time-averaged Poynting vector calculated using Lumerical FDTD commercial mode solving software, and $n_2$ $(x, y)$ is the Kerr coefficient of the different material regions. These calculations assumed picosecond optical pulses having a relatively small spectral width (< 10 nm). We therefore neglected any variation in $n_2$ arising from its dispersion and used $n_2$ instead of the more general third-order nonlinearity $\chi^{(3)}$ in our subsequent analysis and discussion. The values of $n_2$ for silica and silicon used in our calculation were $2.60 \times 10^{-20}$ m$^2$/W [5] and $6.03 \times 10^{-18}$ m$^2$/W, respectively, the latter obtained by fitting the experimental results for the bare SOI nanowire. Note that $\gamma$ in **Eq. (10)** is an effective nonlinear parameter weighted not only by $n_2$ $(x, y)$ but also by $n_0$ $(x, y)$ in the different material regions, which is more accurate for high-index-contrast hybrid waveguides studied here as compared with the theory in Refs. [58, 59].

**Figure 9(a)** shows the ratio of the power in GO on both sidewalls to that in all of the GO material regions caculated by FDTD simulations. The ratio is < 3% for all devices (i.e., all numbers of GO layers), indicating that the TE mode overlap with GO on the sidewalls was negligible compared with the top of the nanowires. We therefore neglected any possible difference between the in-plane and out-of-plane $n_2$ of the layered GO films.

**Figure 9(b)** shows $n_2$ of the GO films versus layer number for fixed coupled pulse energies of 8.2 pJ and 51.5 pJ. The values of $n_2$ are over 3 orders of magnitude higher than that of silicon and agree reasonably well with our previous waveguide FWM [43] and Z-scan measurements [42]. Note that the layer-by-layer characterization of $n_2$ for GO is challenging for Z-scan measurements due to the weak response of extremely thin 2D films [40, 42]. The high $n_2$ of GO films highlights their strong Kerr nonlinearity for not only SPM but also other third-order ($\chi^{(3)}$) nonlinear processes such as FWM, and possibly even enhancing $\chi^{(3)}$ for third harmonic generation (THG) and stimulated Raman scattering, parametric gain, quantum optics, and other processes [60-70]. In **Figure 9(b)**, $n_2$ (both at 51.5 pJ and 8.2 pJ) decreases with GO layer number, showing a similar trend to WS$_2$ measured by a spatial-light system [71]. This is probably due to increased inhomogeneous defects within the GO layers as well as imperfect contact between the different GO layers. At 51.5 pJ, $n_2$ is slightly higher than at 8.2 pJ, indicating a more significant change in the GO optical properties with inceasing power. We also note that the decrease in $n_2$ with GO layer number becomes more gradual for thicker GO films, possibly reflecting the transition of the GO film properties towards bulk material properties − with a thickness independent $n_2$.

**Figure 9(c)** shows the extracted $n_2$ of the GO films versus the coupled pulse energy. In contrast with the monotonic decrease in $n_2$ with GO layer number, $n_2$ varies non-monotonically with the coupled pulse energy. Note that the variation of $n_2$ is much lower than in our previous FWM experiments using CW light [44], probably due to weaker photo thermal effects induced by picosecond optical pulses having much lower average powers compared to CW light. Since it was difficult to accurately measure the slight change in film thickness and refractive index of GO with the coupled pulse energy during the SPM, we neglected any changes in these parameters in our calculations. In principle, this approximation could lead to slight deviations in $n_2$, possibly explaining the non-monotonic relationship between $n_2$ and the coupled pulse energy. Despite this, the fit $n_2$ can still be regarded as a parameter reflecting the overall Kerr nonlinear performance at different coupled pulse energies.

After our SPM experiment, we used low-power (0 dBm) CW light to retest the insertion loss of the GO-coated SOI nanowires. No significant change was observed, and the measured output SPM spectra in **Figure 6** were repeatable when we reinjected the optical pulses, indicating that the optically induced changes (e.g., loss, $n_2$) of the GO films were not permanent. Note that we previously showed that the material properties of GO can also be permanently changed by femtosecond laser pulses [38-41] with significantly higher peak powers. This is distinct from the reversible power-dependent changes observed here.

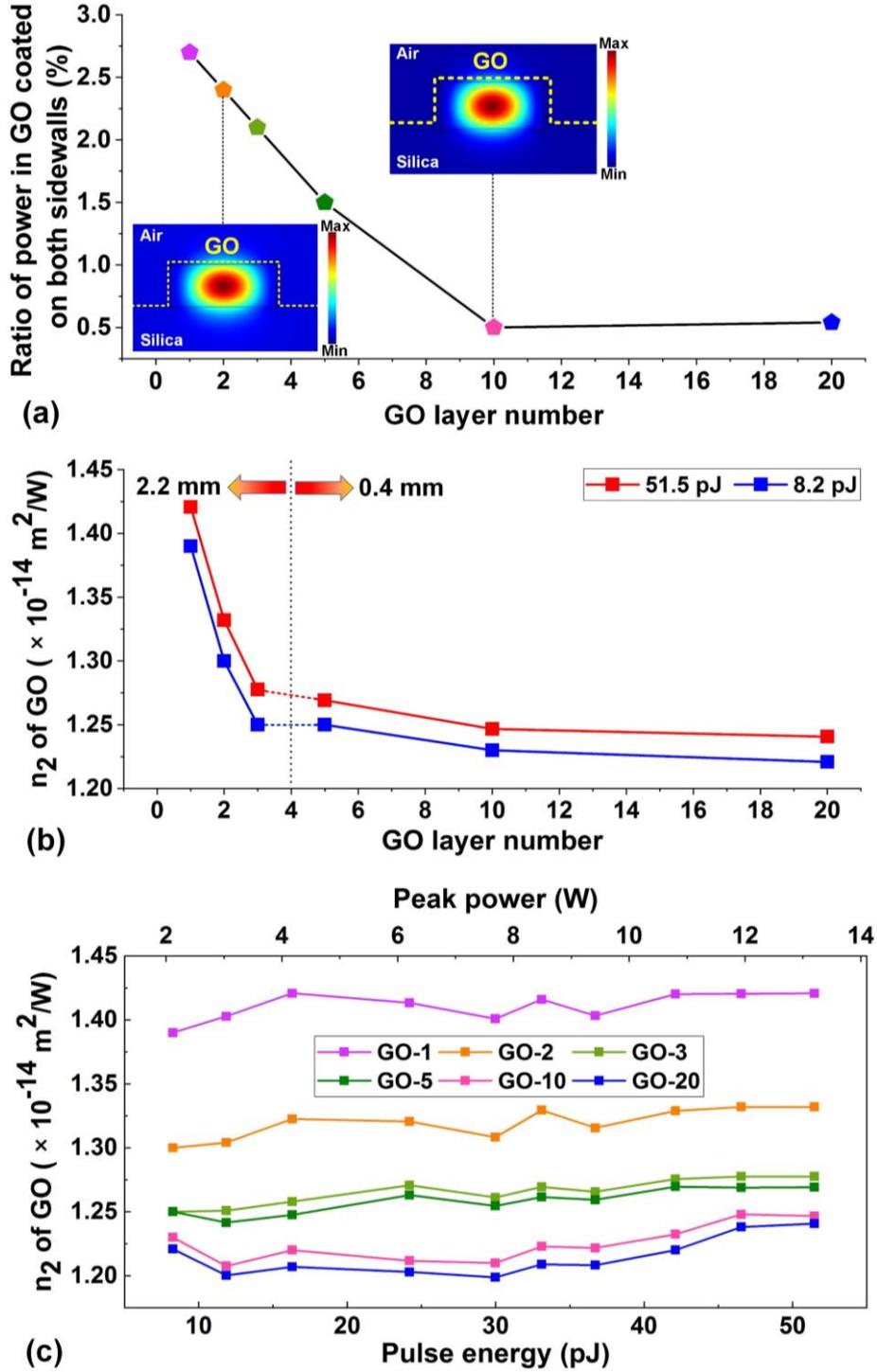

**Figure 9**. (a) Ratio of power in GO coated on both sidewalls to that in all GO material regions. Insets show the TE mode profiles for the SOI nanowires with 2 and 10 layers of GO, respectively. (b) $n_2$ of GO versus layer number at fixed coupled pulse energies of 8.2 pJ and 51.5 pJ. (c) $n_2$ of GO versus coupled pulse energy (or coupled peak power) for the SOI nanowires with (i) 2.2-mm-long, 1−3 layers of GO and (ii) 0.4-mm-long, 5−20 layers of GO.

The layer-number and power-dependent material properties of the layered GO films enable many new device features that are difficult to achieve for silicon photonic devices. We believe this could enable one to tailor the device performance for diverse applications beyond enhancing the Kerr nonlineary as reported here.

*4.3 FOM of the hybrid waveguides*

To quantitively analyze the improvement in the nonlinear performance of the GO-coated SOI nanowires, we calculated the effective nonlinear FOM defined as [5, 6]:

$$FOM_{eff} = \frac{n_{2,\,eff}}{\lambda_c \, \beta_{TPA,\,eff}} \qquad (11)$$

where $\beta_{TPA,\,eff}$ is the effective TPA coefficient obtained by fitting the results in **Figure 4(b)** and $n_{2,\,eff}$ is the effective Kerr coefficient calculated by [6, 27]:

$$n_{2,eff} = \frac{\lambda_c \gamma_{eff} A_{eff}}{2\pi} \qquad (12)$$

**Figure 10(a)** shows the effective Kerr coefficient $n_{2,\,eff}$ and effective TPA coefficient $\beta_{TPA,\,eff}$ (nomalized to those of silicon) of the hybrid waveguides with different numbers of GO layers, where we see that $n_{2,eff}$ increases with GO layer number. In particular, for the hybrid waveguides with 10 or more GO layers, $n_{2,eff}$ is an order of magnitude higher than the bare SOI nanowire. Further, at the same time $\beta_{TPA,\,eff}$ slightly decreases with GO layer number as a result of the influence of the SA in the GO layers. The resulting *FOM_{eff}* (normalized to the FOM of silicon) is shown in **Figure 10(b)** where we see that a very high *FOM_{eff}* of 19 times that of silicon is achieved for the hybrid SOI nanowires with 20 layers of GO. This is remarkable since it indicates that by coating SOI nanowires with GO films, not only can the nonlinearity be significantly enhanced but the relative effect of nonlinear absorption can be greatly reduced as well. This is interesting given that the GO films themselves cannot be described by a nonlinear FOM since the nonlinear absorption displays SA rather than TPA, and yet nonetheless the GO films still are able to reduce the $\beta_{TPA,\,eff}$ of the hybrid waveguides, thus improving the overall nonlinear performance.

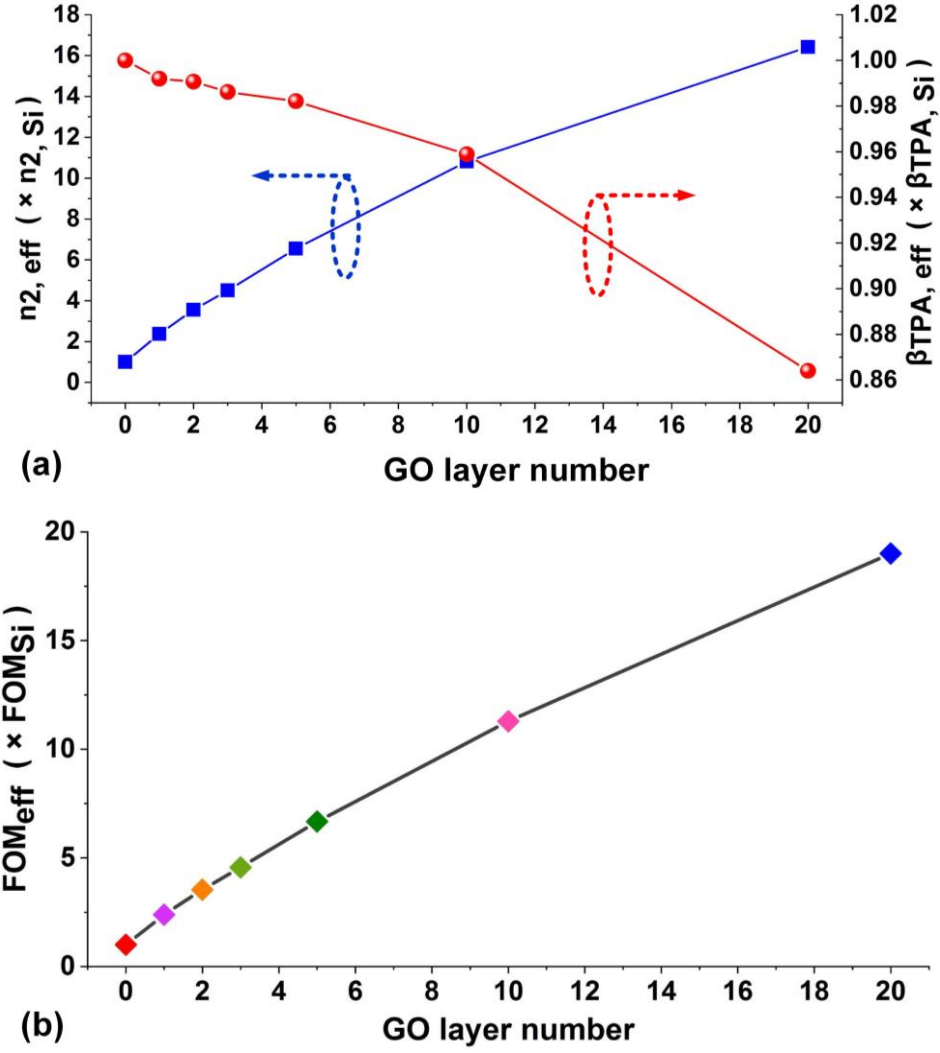

**Figure 10.** (a) $n_{2,\,eff}$ and $\beta_{TPA,\,eff}$ of the hybrid waveguides versus GO layer number. (b) $FOM_{eff}$ of the hybrid waveguides versus GO layer number. The coupled pulse energy in (a)−(b) is ~51.5 pJ. In (a) and (b), the corresponding results for silicon (GO layer number = 0) are also shown for comparison.

Aside from the Kerr nonlinearity ($n_2$, $\gamma$) and the nonlinear FOM ($FOM_{eff}$), the other remaining key factor affecting nonlinear performance is the linear loss. As mentioned the loss of our hybrid waveguides is about 2 orders of magnitude lower than comparable devices integrated with graphene [48], and this played a key role for achieving high BFs in our SPM experiments. Nonetheless, the linear loss did pose a limitation for our devices – otherwise the maximum performance would have been achieved for the thickest films where $\gamma_{eff}$ was the greatest. Reducing the linear loss of the GO films further will significantly improve our device performance even more. We note that the linear loss of the GO films is not a fundamental property − in theory, GO films with a bandgap > 2 eV should have negligible linear loss in spectral regions below the bandgap. Therefore, by optimizing our GO synthesis and coating processes, such as using GO solutions with reduced flake sizes and increased purity, it is anticipated that the loss of our GO films can be significantly reduced.

Finally, we contrast these results with our previous demonstration of enhanced FWM in doped silica devices integrated with layered GO films [43, 44]. In that work the waveguides were very different, having negligible nonlinear absorption and comparatively weak nonlinearity, and so it was perhaps not surprising that the GO layers had a dramatic impact on the nonlinear performance. In contrast, silicon already has a very large Kerr nonlinearity − about twenty times higher than doped silica – and so it is perhaps not obvious that integrating 2D layered GO films would significantly enhance the effective nonlinearity. Even more remarkably, by coating SOI nanowires with GO, the very low nonlinear FOM of silicon of about 0.3, which has posed significant limitations for its nonlinear optical performance [6], can be dramatically improved by almost 20 times. Thus, coating SOI nanowires with GO films can effectively transform silicon into a viable and highly performing nonlinear optical platform.

This is can be attributed to the ultrahigh Kerr nonlinearity of the GO films (about 4 orders of magnitude higher than silicon) as well as the significantly enhanced mode overlap between GO and the SOI nanowires, having much smaller dimensions (500 nm × 220 nm in contrast to 2 μm × 1.5 μm for doped silica waveguides). Mode overlap is important for optimizing the trade-off between Kerr nonlinearity enhancement and loss increase when introducing 2D layered GO films onto different integrated platforms to enhance the nonlinear optical performance. By redesigning the waveguide cross section to optimize the mode overlap, the nonlinear performance such as SPM spectral broadening and FWM conversion efficiency can be further improved. Considering the wide use of SOI devices for diverse applications based on the Kerr nonlinearity, GO-silicon hybrid devices that provide a significantly improved Kerr nonlinearity and nonlinear FOM could play an important role in nonlinear integrated photonic devices well beyond the SPM spectral broadening enhancement reported here.

## 5. Conclusion

We demonstrate an enhanced Kerr nonlinearity in SOI nanowires integrated with 2D layered GO films. The GO films are integrated onto CMOS-compatible SOI nanowires with different lengths over opened windows based on a large-area, transfer-free, layer-by-layer GO coating method. This yields precise control of the film thickness, placement and coating length. We perform detailed SPM measurements in the SOI nanowires with GO films and achieve significant spectral broadening enhancement for the hybrid waveguides. By fitting the experimental results with theory, we obtained the change of GO's third-order nonlinearity with layer number and pulse energy. We observe an enhancement in the nonlinear parameter of up to 16 times and an improved nonlinear FOM of up to a factor of 19. These results highlight the extremely powerful approach of incorporating layered GO films into SOI nanowires to greatly enhance their Kerr nonlinear optical performance, thus transforming silicon into a highly performing nonlinear optical platform.

**Acknowledgements**

Y. Zhang and J. Wu contribute equally to this work. This work was supported by the Australian Research Council Discovery Projects Programs (No. DP150102972 and DP190103186), the Swinburne ECR-SUPRA program, the Industrial Transformation Training Centers scheme (Grant No. IC180100005), and the Beijing Natural Science Foundation (No. Z180007). The authors also acknowledge the Swinburne Nano Lab for the support in device fabrication and characterization. RM acknowledges support by the Natural Sciences and Engineering Research Council of Canada (NSERC) through the Strategic, Discovery and Acceleration Grants Schemes, by the MESI PSR-SIIRI Initiative in Quebec, and by the Canada Research Chair Program. He also acknowledges additional support by the Professorship Program (grant 074-U 01) and by the 1000 Talents Sichuan Program in China.

<div style="text-align: right">Received: (will be filled in by the editorial staff)<br>Revised: (will be filled in by the editorial staff)<br>Published online: (will be filled in by the editorial s</div>